\documentclass[11pt,superscriptaddress,aps,prd,preprint]{revtex4}
\usepackage{amsmath}

\makeatletter
\usepackage[T1]{fontenc}
\usepackage{amsmath}
\usepackage{amssymb}
\usepackage{graphicx}
\usepackage{tabularx}

\usepackage{slashed}

\newcommand{\bea}{\begin{eqnarray}}
\newcommand{\eea}{\end{eqnarray}}

\begin{document}

%\title{Scattering in Phase Space}
\title{Stefan-Boltzmann  law and Casimir effect for the scalar field in phase space at finite temperature}

\author{R. G. G. Amorim}\email[]{ronniamorim@gmail.com}
 \affiliation{Faculdade Gama, Universidade de Bras\'{i}lia, Setor Leste
(Gama), 72444-240, Bras\'{i}lia-DF, Brazil.}

\author{J. S da Cruz Filho}\email[]{ze_filho@fisica.ufmt.br}
\affiliation{Instituto de F\'{\i}sica, Universidade Federal de Mato Grosso,\\
78060-900, Cuiab\'{a}, Mato Grosso, Brazil}

\author{A. F. Santos}\email[]{alesandroferreira@fisica.ufmt.br}
\affiliation{Instituto de F\'{\i}sica, Universidade Federal de Mato Grosso,\\
78060-900, Cuiab\'{a}, Mato Grosso, Brazil}

\author{S. C. Ulhoa}\email[]{sc.ulhoa@gmail.com}
\affiliation{Instituto de F\'{i}sica,
Universidade de Bras\'{i}lia, 70910-900, Bras\'{i}lia, DF,
Brazil.}

\begin{abstract}
In this paper we use the scalar field constructed in phase space to analyze the analogous Stefan-Boltzman law and Casimir effect both of them at finite temperature. The temperature is introduced by Thermo Field Dynamics (TFD) and the quantities are analyzed once projected in the coordinates space.

\end{abstract}

\maketitle

\section{Introduction}
Eugene Paul Wigner \cite{Wigner,Hillery} introduced in 1932 the first formalism to quantum mechanics in phase space,
motived by finding a way to treat transport equations for superfluids.
Wigner formalism allows  mapping between  quantum operators, say  A,  defined in the Hilbert
space, ${\cal S}$, with classical functions,  say $a_w(q,p)$, in phase space $\Gamma$, through the $\ast $ is the
star-product or Moyal-product. The opposite problem, i. e., from a classical function finding the correspondent operator, is accomplished by the Weyl mapping. The main motivation to define a given physical theory in phase space is its natural interpretation as demonstrated by classical mechanics.

The star-product has been explored in phase space in different ways. Particularly, it has been used
to define operators like $a_{w}(q,p)\ast $ of interest to study irreducible unitary representations of
kinematical groups in phase space \cite{Oliveira}. In case of non-relativistic symmetries, this leads to a
Schr\"{o}dinger   equation in phase space, where the wave function is directly associated with
the Wigner function, so with full physical meaning. In this formalism of quantum mechanic,
the observables are represented by operators of type
$\widehat{a}=a_{w}\star $, which are used to construct a representation of Galilei symmetries.
The Wigner function is given by $f_{w}(q,p)=\psi ^{\dag }\star \psi $ where $\psi =\psi (q,p)$
are the wave functions, solutions of the Schr\"{o}dinger  equation represented in
phase space. Since it is a theory of representation, this formalism has been generalized to the
relativistic case, leading to Klein-Gordon and Dirac equations in phase space \cite{Ronni}.
This method has been applied successfully, for instance, to the  analysis of abelian
gauge symmetries \cite{Ronni01}, describing the dynamics
(interaction) in the formulation of quantum theory in phase space.

Although the quantum field theory has successfully been applied to several systems, it doesn't take into account the temperature of such systems. This is a fundamental problem since all macroscopic features of a quantum model is related to temperature. Among all schemes to introduce temperature in a physical theory we cite two ways. The first one is to interpret time as temperature by a Wick rotation \cite{wick}. This approach is problematic when one is dealing with a time evolution of a physical state. The second one the so called Thermo Field Dynamics (TFD), it's a natural way to deal with dynamical systems. It preserves the time-evolution once the temperature is identified with a rotation in a duplicated Fock space \cite{khanna}. In this article we explore how to implement TFD in phase space, particularly we analyze the scalar field in phase space at finite temperature. The Casimir effect for scalar field in phase space at zero and finite temperature is calculated.

The Casimir effect \cite{Casimir} is measured when two parallel conducting plates are attracted due to vacuum fluctuations. Although the first application has been developed for the electromagnetic field all quantum field should exhibit this phenomenon. In fact it was demonstrated that non-relativistic fields such as in Schrödinger equation the Casimir effect is present \cite{artigokhanna}. Particularlly for non-relativistic fields such a effect is physical only at finite temperature. Sparnaay \citep{Sparnaay} made the first experimental observation of the Casimir effect. Subsequent experiments have established this effect to a high degree of accuracy \cite{Lamoreaux, Mohideen}.

The article is divided as follows. In section \ref{kg}, the symplectic Klein-Gordon field is introduced. 
In section \ref{cqsf}, we show the canonical quantization of scalar field. 
Then in section \ref{tfd}, we recall the ideas of Thermo Field Dynamics. In section \ref{em}, we calculate the energy-momentum tensor of the symplectic scalar field and in section \ref{applications}, we derive the Stefan-Boltzmann like law and the Casimir effect at finite temperature. Finally in the last section se present our conclusions.

\section{Symplectic Klein-Gordon Field and Wigner Function}\label{kg}

Let us define the star-operator as
\begin{align*}
\widehat{A}& =A(q,p)\star =A(q,p)\exp \left[ \frac{i\hbar }{2}(\frac{%
\overleftarrow{\partial }}{\partial q}\frac{\overrightarrow{\partial }}{%
\partial p}-\frac{\overleftarrow{\partial }}{\partial p}\frac{%
\overrightarrow{\partial }}{\partial q})\right] \\
& =A(q+\frac{i\hbar }{2}\partial _{p},p-\frac{i\hbar }{2}\partial _{p}),
\end{align*}%

hence we can derive the following operators
\begin{equation}\label{op1}
\widehat{P}^{\mu}=p^{\mu}\star=
p^{\mu}-\frac{i}{2}\frac{\partial}{\partial q_{\mu}},
\end{equation}

\begin{equation}\label{op2}
\widehat{Q}^{\mu}=q^{\mu}\star =
q^{\mu}+\frac{i}{2}\frac{\partial}{\partial p_{\mu}},
\end{equation}
 and
\begin{equation}\label{op3}
\widehat{M}_{\nu\sigma}=M_{\nu\sigma}\star
=\widehat{Q}_{\nu}\widehat{P}_{\sigma}-\widehat{Q}_{\sigma}\widehat{P}_{\nu}.
\end{equation}

These operators satisfy the Poincaré algebra and act in Hilbert space associated with phase space $\mathcal{H}(\Gamma)$.
From them, we construct a symplectic representation of Poincaré-Lie algebra and as a result we obtain the Klein-Gordon equation in phase space \cite{Ronni}
\begin{equation}\label{kg1}
\widehat{P}^{2}\phi(q,p)=p^2\star\phi(q,p)=m^2\phi(q,p)
\end{equation}
\begin{equation}\label{kg2}
(p^{\mu }p_{\mu }-ip^{\mu}\frac{\partial }{\partial q^{\mu }}-\frac{1}{4}%
\frac{\partial }{\partial q^{\mu }}\frac{\partial }{\partial q_{\mu }}%
)\phi(q,p)=m^2\phi(q,p).
\end{equation}
The functions $\phi(q,p)$ are defined in phase space $\Gamma$ and satisfy the condition
$$\int d^{4}qd^{4}p\phi^{\dagger}(q,p)\phi(q,p)<\infty.$$
Eq.(\ref{kg2}) can be derived from lagrangian given by \cite{Ronni01}
\begin{eqnarray}
 \mathcal{L}
&=&- \left( D^{\mu}  \star \phi  \right) \star \left( \phi^{\ast} \star D_{\mu} \right)
+ m^2 \phi^{\ast} \star \phi,
 \label{eq2}
\end{eqnarray}
where $D^{\mu}=p^{\mu}-\frac{i}{2}\frac{\partial}{\partial q_{\mu}}$.
The association with Wigner formalism is obtained from
\begin{equation}\label{wigner1}
f_W(q,p)=\phi(q,p)\star\phi^{\ast}(q,p).
\end{equation}.
To show this, we multiply the right-hand side of Eq.(\ref{kg1}) by $\phi^{\ast}(q,p)$,
\begin{equation}\label{demo1}
(p^2\star\phi(q,p))\star\phi^{\ast}(q,p)=m^2\phi(q,p)\star\phi^{\ast}(q,p),
\end{equation}
but since $\phi^{\ast}(q,p)\star p^2=m^2\phi^{\ast}(q,p)$, we also have
\begin{equation}\label{demo2}
\phi(q,p)\star(\phi^{\ast}(q,p)\star p^2)=m^2\phi(q,p)\star\phi^{\ast}(q,p).
\end{equation}
Subtracting Eq.(\ref{demo1}) of Eq.(\ref{demo2}), and using the associativity of star-product, we get
\begin{equation}\label{wigner2}
\{p^2,f_W(q,p)\}_{M}=0,
\end{equation}
where the Moyal-bracket is given by
$$\{a,b\}_M=a\star b- b \star a.$$
Calculating, we obtain
\begin{equation}\label{wigner3}
p_{\mu}\frac{\partial f_W(q,p)}{\partial q_{\mu}}=0,
\end{equation}
a well known result. Another properties of Wigner function, such as non-positiveness, can be derived analogous a non-relativistic case \cite{Oliveira}.

If we consider the interaction potential $V$, the follow density of lagrangian should be used

\begin{equation}\label{lag1}
\mathcal{L} = \frac{-1}{4}\frac{\partial\psi}{\partial
q_{\mu}}\frac{\partial\psi^{\dagger}}{\partial q^{\mu}} +
\frac{1}{2}ip^{\mu}(\psi^{\dagger}\frac{\partial\psi}{\partial
q^{\mu}}-\psi \frac{\partial\psi^{\dagger}}{\partial q^{\mu}}) -
(p^{\mu}p_{\mu}-m^{2})\psi\psi^{\dagger}+ U(\psi\psi^{\dagger}).
\end{equation}

This lagrangian induces the equation

\begin{equation}\label{int1}
\frac{-1}{4}\frac{\partial^{2}\psi}{\partial q^{\mu}\partial
q_{\mu}}-ip^{\mu}\frac{\partial\psi}{\partial
q^{\mu}}+(p^{\mu}p_{\mu}-m^{2})\psi-V(\psi)=0,
\end{equation}
where $V(\psi)=\frac{\partial
U(\psi\psi^{\dagger})}{\partial\psi^{\dagger}}$.

Solutions for Eq.(\ref{int1}) can be obtained from Green function method. For this proposal, take the function
$G=G(q^{\mu},q'^{\mu},p^{\mu},p'^{\mu})$ in which satisfy

\begin{equation}\label{int2}
\frac{-1}{4}\frac{\partial^{2}G}{\partial q^{\mu}\partial
q_{\mu}}-ip^{\mu}\frac{\partial G}{\partial
q^{\mu}}+(p^{\mu}p_{\mu}-m^{2})G=\delta(q^{\mu}-q'^{\mu})\delta(p^{\mu}-p'^{\mu}),
\end{equation}
where  $G$ is the Green function. By superposition principle, solution of Eq.(\ref{int2}) is given by

\begin{equation}\label{int3}
\psi(q^{\mu},p^{\mu})= \psi_{0}(q^{\mu},p^{\mu})+\int
d^{4}q'^{\mu}d^{4}p'^{\mu}G(q^{\mu},q'^{\mu},p^{\mu},p'^{\mu})V(\psi),
\end{equation}
where $\psi_{0}(q^{\mu},p^{\mu})$ is the solution of free case.

We can find the solution of Eq.(\ref{int3}) taking its Fourier transform. Defining  $F[\frac{\partial^{2}G}{\partial
q^{\mu}\partial q_{\mu}}]=-k^{2}\widetilde{G}(k^{\mu},\eta^{\mu})$,
$F[\frac{\partial G}{\partial
q^{\mu}}]=-ik^{\mu}\widetilde{G}(k^{\mu},\eta^{\mu})$ and
$F[G]=\widetilde{G}(k^{\mu},\eta^{\mu}),$ follows up

\begin{equation}\label{int4}
\frac{1}{4}k^{2}\widetilde{G}(k^{\mu},p^{\mu}) -
p^{\mu}k_{\mu}\widetilde{G}(k^{\mu},p^{\mu})+
(p^{\mu}p_{\mu}-m^{2})\widetilde{G}(k^{\mu},p^{\mu})=1,
\end{equation}
where  $F[g]$ stands the Fourier transform os function $g$. In this way, we obtain

\begin{equation}\label{int5}
\widetilde{G}(k^{\mu},p^{\mu})=\frac{\delta(p^{\mu}-p'^{\mu})}{\frac{1}{4}k^{2}
- p^{\mu}k_{\mu}+ (p^{\mu}p_{\mu}-m^{2})}.
\end{equation}
Then,

\begin{eqnarray}\label{int6}
G(q^{\mu},q'^{\mu},p^{\mu},p'^{\mu})&=&\frac{1}{(2\pi)^{4}}\int
d^{4}k^{\mu}e^{-ik^{\mu}q_{\mu}}\widetilde{G}(k^{\mu},p^{\mu})\nonumber
\\&=&\frac{1}{(2\pi)^{4}}\int
d^{4}k^{\mu}\frac{\delta(p^{\mu}-p'^{\mu})e^{-ik^{\mu}q_{\mu}}}{\frac{1}{4}k^{2}
- p^{\mu}k_{\mu}+ (p^{\mu}p_{\mu}-m^{2})}.
\end{eqnarray}
The solution is
\begin{equation}
\psi(q^{\mu},p^{\mu})= \psi_{0}(q^{\mu},p^{\mu})+\int
d^{4}q'^{\mu}d^{4}p'^{\mu}G(q^{\mu},q'^{\mu},p^{\mu},p'^{\mu})V(\psi).
\end{equation}

Wigner function can be derived from Green function by \cite{amorim2}

\begin{equation}\label{wigreen}
f_W(q^{\mu},p^{\mu})=\lim_{q^{\mu '}p^{\mu '}\rightarrow q^{\mu}p^{\mu}}\exp{i\left(\frac{\partial}{\partial q^{\mu}}\frac{\partial}{\partial p^{\mu '}}-\frac{\partial}{\partial q^{\mu '}}\frac{\partial}{\partial p^{\mu}}\right)}G(q^{\mu},q'^{\mu},p^{\mu},p'^{\mu}).
\end{equation}

Eq.(\ref{wigreen}) provide a method to describe interaction process and scattering theory with physical interpretation in phase space.
\section{Canonical Quantization of Scalar Field in Phase Space}\label{cqsf}

In this section we construct the formalism of canonical quantization of Klein-Gordon field in phase space. From this formalism we obtain the propagator written in phase space.

From lagrangian given in Eq.(\ref{lag1}) we define the conjugate momenta associated to fields $\phi(q,p)$ e
$\phi^{\dagger}(q,p)$ by

$$\pi=\frac{\partial \mathcal{L}}{(\partial_{0}\phi)}$$

and

$$\pi^{\dagger}=\frac{\partial \mathcal{L}}{(\partial_{0}\phi^{\dagger})}.$$

After some calculations we have

$$\pi= \frac{-1}{4}\frac{\partial \phi^{\dagger}}{\partial
q^{0}}+\frac{1}{2}ip_{0}\phi^{\dagger}=\frac{-1}{2}ip^{0}\star\phi^{\dagger}$$

and

$$\pi^{\dagger}= \frac{1}{4}\frac{\partial \phi}{\partial
q^{0}}-\frac{1}{2}ip_{0}\phi=\frac{1}{2}ip^{0}\star\phi.$$

Follow usual procedure of quantization, we impose the commutation relations

\begin{equation}
[\pi(q,q_{0};p),\phi(q\prime,q_{0};p\prime)]=i
\delta(q-q\prime)\delta(p-p\prime),\notag
\end{equation}
and
\begin{equation}
[\pi^{\dagger}(q,q_{0};p),\phi^{\dagger}(q\prime,q_{0};p\prime)]=i
\delta(q-q\prime)\delta(p-p\prime),\notag
\end{equation}
the other commutation relations are nulls.

\subsection{Annihilation and Creation Operators}

The fields $\phi(q,p)$ and $\phi^{\dagger}(q,p)$ may be expanded as
\begin{equation}\label{exp1}
\phi(q,p)=\int
\frac{d^{3}k}{[(2\pi)^{3}2\omega_{k}]^{1/2}}[a(k,p)e^{-ikq} +
b^{\dagger}(k,p)e^{ikq}]
\end{equation}
and
\begin{equation}\label{exp2}
\phi^{\dagger}(q,p)=\int
\frac{d^{3}k}{[(2\pi)^{3}2\omega_{k}]^{1/2}}[b(k,p)e^{-ikq} +
a^{\dagger}(k,p)e^{ikq}],
\end{equation}
where
$\omega_{k}=[(\frac{1}{2}\mathbf{k}-\mathbf{p})^{2}+m^{2}]^{1/2}$ and the canonical variable related to position is $K$, i.e.,
$q\rightarrow k$.

The functions

$$f_{k}(q)= \frac{e^{-i(kq}}{[(2\pi)^{3}2\omega_{k}]^{\frac{1}{2}}},$$
form a ortonormal set
\begin{equation}
\int
f^{\ast}_{k}(q)\widetilde{p^{0}\star}f_{k^{\prime}}(q)d^{3}q=\delta^{3}(\mathbf{k}-\mathbf{k}^{\prime}),
\nonumber
\end{equation}
where $A\widetilde{p^{0}\star}B= A (p^{0}\star B) - (p^{0}\star
A)B$.

In this way, the fields $\phi(q,p)$ and $\phi^{\dagger}(q,p)$ may be written as 
\begin{equation}\label{field1}
\phi(q,p)=\int
\frac{d^{3}k}{[(2\pi)^{3}2\omega_{k}]^{1/2}}[a(k,p)f_{k}(q) +
b^{\dagger}(k,p)f^{\ast}_{k}(q)],
\end{equation}
and
\begin{equation}\label{field2}
\phi^{\dagger}(q,p)=\int
\frac{d^{3}k}{[(2\pi)^{3}2\omega_{k}]^{1/2}}[b(k,p)f_{k}(q) +
a^{\dagger}(k,p)f^{\ast}_{k}(q)].
\end{equation}
Inverting Eq.(\ref{field1}) and Eq.(\ref{field2}) we obtain
\begin{equation}\label{op1}
a(k,p)=\int d^{3}q
[(2\pi)^{3}2\omega_{k}]^{\frac{1}{2}}f^{\ast}_{k}\widetilde{p^{0}\star}\phi(q,p);\nonumber
\end{equation}
\begin{equation}\label{op2}
b(k,p)=\int d^{3}q
[(2\pi)^{3}2\omega_{k}]^{\frac{1}{2}}f^{\ast}_{k}\widetilde{p^{0}\star}\phi^{\dagger}(q,p);\nonumber
\end{equation}
\begin{equation}\label{op3}
a^{\dagger}(k,p)=\int d^{3}q
[(2\pi)^{3}2\omega_{k}]^{\frac{1}{2}}\phi^{\dagger}(q,p)\widetilde{p^{0}\star}f_{k};\nonumber
\end{equation}
\begin{equation}\label{op4}
b^{\dagger}(k,p)=\int d^{3}q
[(2\pi)^{3}2\omega_{k}]^{\frac{1}{2}}\phi(q,p)\widetilde{p^{0}\star}f_{k}.\nonumber
\end{equation}
We can show that

\begin{equation}\label{rela1}
[a(k,p),a^{\dagger}(k^{\prime},p^{\prime})]=(2\pi)^{3}2\omega_{k}\delta(\mathbf{k}-\mathbf{k^{\prime}}),
\end{equation}

and
\begin{equation}\label{rela2}
[b(k,p),b^{\dagger}(k,p)]=(2\pi)^{3}2\omega_{k}\delta(\mathbf{k}-\mathbf{k^{\prime}}).
\end{equation}

The operators $a(k,p)$, $a^{\dagger}(k,p)$, $b(k,p)$ and
$b^{\dagger}(k,p)$ play a crucial role in the particle interpretation  of the quantised field theory. First, define the operators
\begin{equation}
N(k,p)=a^{\dagger}(k,p)a(k,p)\nonumber
\end{equation}
and
\begin{equation}
M(k,p)=b^{\dagger}(k,p)b(k,p).\nonumber
\end{equation}
It is simple to show that $N(k,p)$ and
$N(k^{\prime},p^{\prime})$ commute
$$[N(k,p),N(k^{\prime},p^{\prime})]=0.$$
In analogous sense,
$$[M(k,p),M(k^{\prime},p^{\prime})]=0.$$
In this case, the eigenstates of these operators may be used to form a basis. Let denote the eigenvalue of $N(k,p)$ by $n(k,p)$ and the eigenvalue of $M(k,p)$ by $m(k,p)$, i.e.,
\begin{equation}
N(k,p)|n(k,p)\rangle=n(k,p)|n(k,p)\rangle,\nonumber
\end{equation}
\begin{equation}
M(k,p)|m(k,p)\rangle=m(k,p)|m(k,p)\rangle.\nonumber
\end{equation}
And now use the commutations relations
$[N(k,p),a^{\dagger}(k,p)]=a^{\dagger}(k,p)$,
$[N(k,p),a(k,p)]=-a(k,p)$, $[M(k,p),b^{\dagger}(k,p)]=b^{\dagger}(k,p)$,
$[M(k,p),b(k,p)]=-b(k,p)$, we find
\begin{equation}\label{cr1}
N(k,p)a^{\dagger}(k,p)|n(k,p)\rangle=(n(k,p)+1)a^{\dagger}(k,p)|n(k,p)\rangle,
\end{equation}
\begin{equation}\label{cr2}
N(k,p)a(k,p)|n(k,p)\rangle=(n(k,p)-1)a(k,p)|n(k,p)\rangle,
\end{equation}
\begin{equation}\label{cr3}
M(k,p)b^{\dagger}(k,p)|m(k,p)\rangle=(m(k,p)+1)b^{\dagger}(k,p)|m(k,p)\rangle,
\end{equation}
\begin{equation}\label{cr4}
M(k,p)b(k,p)|m(k,p)\rangle=(m(k,p)-1)b(k,p)|m(k,p)\rangle.
\end{equation}
Eqs. (\ref{cr1},\ref{cr2}) tell us that if the state $|k,p\rangle$
has eigenvalue $n(k,p)$,and the states
$a^{\dagger}(k,p)|n(k,p)\rangle$ and $a(k,p)|n(k,p)\rangle$ are eigenstates  of $N(k,p)$ with respective eigenvalues $n(k,p)+1$ and $n(k,p)-1$. And analogous, we note  by Eqs. (\ref{cr3},\ref{cr4}) tell us that if the state $|k,p\rangle$
has eigenvalue $m(k,p)$, and the states
$b^{\dagger}(k,p)|m(k,p)\rangle$ and $b(k,p)|m(k,p)\rangle$ are eigenstates  of $M(k,p)$ with respective eigenvalues $m(k,p)+1$ and $m(k,p)-1$. So, the operators $a^{\dagger}(k,p)$ and
$a(k,p)$ are interpreted as creation and annihilation operators of particles, respectively. Then, analogously,
$b^{\dagger}(k,p)$ and $b(k,p)$ can be interpreted as
creation and annihilation operators of antiparticles, respectively.

Using the creation and annihilation operators, the Hamiltonian of scalar field in phase space can be written by
\begin{equation}
H=\int \frac{d^{3}k}{(2\pi)^{3}2\omega_{k}}[a^{\dagger}(k,p)a(k,p) +
b^{\dagger}(k,p)b(k,p)].\nonumber
\end{equation}

We can also show that the particles that are the quantum of the Klein-Gordon field
obey the Bose-Einstein statistics. For this, note that 
basta notarmos que
\begin{equation}\nonumber
|n(k,p),m(k,p)\rangle=\frac{1}{(n(k,p)!m(k,p)!)^{1/2}}[a^{\dagger}(k,p)]^{n}[b^{\dagger}(k,p)]^{m}|0\rangle.
\end{equation}

The connection between the solution of free Klein-Gordon equation, $\varphi(q,p)=\xi(p^{\mu})e^{-ik_{\mu}q^{\mu}}$,  and the canonical quantization formalism is given by
\begin{equation}
\varphi(q,p)=\langle 0|\phi(q,p)|k,p\rangle.
\end{equation}

In sequence, we establish the association between Green function given in Eq.(\ref{int6}) and the expression $\langle 0|T[\phi
(q,p)\phi ^{\ast }(q^{\prime },p^{\prime })]|0\rangle$. For this propose, we consider the Green function
\begin{equation}
G(q^{\mu},p^{\mu},p'^{\mu})=\frac{1}{(2\pi)^{4}}\int
d^{4}k^{\mu}\frac{\delta(p^{\mu}-p'^{\mu})e^{-ik^{\mu}q_{\mu}}}{(\frac{1}{2}k_{0}-p_{0})^{2}-[(\frac{1}{2}\mathbf{k}-\mathbf{p})^{2}+m^{2}]},
\end{equation}
in which can be written in the form
\begin{equation}
G(q^{\mu},p^{\mu},p'^{\mu})=\delta(p-p^{\prime})\int\frac{d^{3}\mathbf{k}dk_{0}}{(2\pi)^{4}}\frac{e^{-ikq}}{2\omega_{k}}\left(\frac{1}{\frac{1}{2}k_{0}-p_{0}-\omega_{k}+i\epsilon}-\frac{1}{\frac{1}{2}k_{0}-p_{0}+\omega_{k}-i\epsilon}\right),
\end{equation}
where $\omega_{k}^{2}=(\frac{1}{2}\mathbf{k}-\mathbf{p})^{2}$. Using Cauchy Theorem we have
\begin{equation}\label{gr1}
G(q^{\mu},q^{\prime\mu},p^{\mu},p'^{\mu})=-2i\frac{\delta(p^{\mu}-p^{\prime\mu})}{(2\pi)^{3}}\int\frac{d^{3}\mathbf{k}}{2\omega_{k}}[\theta(q_{0}-q^{\prime}_{0})e^{-ik(q_{0}-q^{\prime}_{0})}+
\theta(q^{\prime}_{0}-q_{0})e^{ik(q_{0}-q^{\prime}_{0})}].
\end{equation}
In order, substituting Eqs.(\ref{field1}, \ref{field2}) in $\langle 0|T[\phi
(q,p)\phi ^{\ast }(q^{\prime },p^{\prime })]|0\rangle$ we obtain
\begin{eqnarray}
\langle 0|T[\phi (q,p)\phi ^{\ast }(q^{\prime },p^{\prime
})]|0\rangle=&&\int\frac{d^{3}\mathbf{k}d^{3}\mathbf{k}^{\prime}}{(2\pi)^{6}(2\omega_{k^{\prime}}\omega_{k})^{1/2}}[\theta(q_{0}-q^{\prime}_{0})e^{-i(kq-k^{\prime}q^{\prime})}\langle
0|a(k,p)a^{\dagger}(k^{\prime},p^{\prime})|0\rangle\nonumber
\\&&+\theta(q^{\prime}_{0}-q_{o})e^{-i(kq^{\prime}-k^{\prime}q)}\langle
0|a(k,p)a^{\dagger}(k^{\prime},p^{\prime})|0\rangle],\nonumber
\end{eqnarray}
where the Heaviside function $\theta(x)$ is defined as $\theta(x)=1$ for $x>1$ and $\theta(x)=0$ for $x<1$. Then, using Eq.(\ref{rela1},\ref{rela2}) we obtain
\begin{equation}
\langle 0|T[\phi (q,p)\phi ^{\ast }(q^{\prime },p^{\prime
})]|0\rangle=\frac{\delta(p^{\mu}-p^{\prime\mu})}{(2\pi)^{3}}\int\frac{d^{3}\mathbf{k}}{2\omega_{k}}[\theta(q_{0}-q^{\prime}_{0})e^{-ik(q_{0}-q^{\prime}_{0})}+
\theta(q^{\prime}_{0}-q_{0})e^{ik(q_{0}-q^{\prime}_{0})}].
\end{equation}
We then have
\begin{equation}\label{prop}
\langle 0|T[\phi
(q,p)\phi ^{\ast }(q^{\prime },p^{\prime
})]|0\rangle=\frac{i}{2}G(q^{\mu},q^{\prime\mu},p^{\mu},p'^{\mu}).
\end{equation}
Eq.({\ref{prop}) show us the connection between the propagator $ \langle 0|T[\phi
(q,p)\phi ^{\ast }(q^{\prime },p^{\prime
})]|0\rangle$ and Green function in phase space.

\section{Thermo Field Dynamics - TFD}\label{tfd}

In this section a brief introduction to TFD formalism is presented \cite{Umezawa1, Umezawa2, Umezawa22, Khanna1, Khanna2}. In this formalism the thermal average of an observable is given by the vacuum expectation value in an extended Fock space , i.e., $\langle A \rangle=\langle 0(\beta)| A|0(\beta) \rangle$, where $|0(\beta) \rangle$ is the thermal vacuum. The foundation of the TFD consist in two ingredients: (i) doubling of the original Hilbert space, i.e., the original Hilbert space ${\cal S}$ of the system is doubled leading to an expanded space ${\cal S}_T={\cal S}\otimes \tilde{\cal S}$ and (ii) Bogoliubov transformations. This doubling is defined by a mapping ($^\thicksim$): ${\cal S}\rightarrow {\tilde{\cal S}}$ associating each operator say $a$, in ${\cal S}$ to two operators in ${\cal S}_T$, such as
\bea
A=a\otimes 1,\quad\quad\quad\quad \tilde{A}=1\otimes a.
\eea
The standard doublet notation for an arbitrary bosonic operator ${\cal X}$ is
\bea
{\cal X}^a&=&\left( \begin{array}{cc} {\cal X}^1\\
{\cal X}^2 \end{array} \right)
=\left( \begin{array}{cc} {\cal X}\\
-\tilde{{\cal X}}^\dagger \end{array} \right),
\eea
where $a,b=1,2$. The physical variables are described by nontilde operators.

The Bogoliubov transformation introduces a rotation in the tilde and nontilde variables. Then the thermal effects are introduced by a Bogoliubov transformation, ${\cal U}(\alpha)$, that is defined as
\bea
{\cal U}(\alpha)=\left( \begin{array}{cc} u(\alpha) & -v(\alpha) \\
-v(\alpha) & u(\alpha) \end{array} \right),
\eea
where $u^2(\alpha)-v^2(\alpha)=1$. These quantities $u(\alpha)$ and $v(\alpha)$ are related to the Bose distribution. The parameter $\alpha$ is associated with temperature, but, in general, it may be associated with other physical quantities. The  Bogoliubov transformations for bosons and fermions are different. For bosons are given as
\bea
a(k)&=&c_B(\omega)a(k,\beta)+d_B(\omega)\tilde a^\dagger(k,\beta),\nonumber\\
a^\dagger(k)&=&c_B(\omega)a^\dagger(k,\beta)+d_B(\omega)\tilde a(k,\beta),\nonumber\\
\tilde a(k)&=&c_B(\omega)\tilde a(k,\beta)+d_B(\omega) a^\dagger(k,\beta),\nonumber\\
\tilde a^\dagger(k)&=&c_B(\omega)\tilde a^\dagger(k,\beta)+d_B(\omega) a(k,\beta),\label{BTp}
\eea
where $(a^\dagger, \tilde{a}^\dagger)$ are creation operators and $(a, \tilde{a})$ are destruction operators, with
\bea
c_B^2(\omega)=1+f_B(\omega),\quad\quad d_B^2(\omega)=f_B(\omega), \quad\quad f_B(\omega)=\frac{1}{e^{\beta\omega}-1},\label{phdef}
\eea
with $\omega=\omega(k)$.

For fermions the Bogoliubov transformations are
\bea
a(k)&=&c_F(\omega)a(k,\beta)+d_F(\omega)\tilde a^\dagger(k,\beta),\nonumber\\
a^\dagger(k)&=&c_F(\omega)a^\dagger(k,\beta)+d_F(\omega)\tilde a(k,\beta),\nonumber\\
\tilde a(k)&=&c_F(\omega)\tilde a(k,\beta)-d_F(\omega) a^\dagger(k,\beta),\nonumber\\
\tilde a^\dagger(k)&=&c_F(\omega)\tilde a^\dagger(k,\beta)-d_F(\omega) a(k,\beta),\label{BTf}
\eea
with
\bea
c_F^2(\omega)=1-f_F(\omega),\quad\quad d_F^2(\omega)=f_F(\omega), \quad\quad f_F(\omega)=\frac{1}{e^{\beta\omega}+1}.\label{ferdef}
\eea

Let's to consider a free scalar field in Minkowski space with $diag(g^{\mu\nu})=(+1,-1,-1,-1)$ and then analyze its propagator. Using the Bogoliubov transformation the $\alpha$-dependent scalar field is given by
\bea
\phi(x;\alpha)&=&{\cal U}(\alpha)\phi(x){\cal U}^{-1}(\alpha).
\eea
There is a similar equation for tilde field. The propagator for the scalar field, $\alpha$-dependent, is
\bea
G_0^{(ab)}(x-x';\alpha)=i\langle 0,\tilde{0}| \tau[\phi^a(x;\alpha)\phi^b(x';\alpha)]| 0,\tilde{0}\rangle,
\eea
where $\tau$ is the time ordering operator. Using $|0(\alpha)\rangle={\cal U}(\alpha)|0,\tilde{0}\rangle$
\bea
G_0^{(ab)}(x-x';\alpha)&=&i\langle 0(\alpha)| \tau[\phi^a(x)\phi^b(x')]| 0(\alpha)\rangle,\nonumber\\
&=&i\int \frac{d^4k}{(2\pi)^4}e^{-ik(x-x')}G_0^{(ab)}(k;\alpha),
\eea
where
\bea
G_0^{(ab)}(k;\alpha)={\cal U}^{-1}(k;\alpha)G_0^{(ab)}(k){\cal U}(k;\alpha),
\eea
with
\bea
{\cal U}(k;\alpha)=\left( \begin{array}{cc} u(k;\alpha) & -v(;,\alpha) \\
-v(k;\alpha) & u(k;\alpha) \end{array} \right), \quad\quad\quad\quad G_0^{(ab)}(k)=\left( \begin{array}{cc} G_0(k) & 0 \\
0 & -G^*_0(k) \end{array} \right),
\eea
and
\bea
G_0(k)=\frac{1}{k^2-m^2+i\epsilon}.
\eea
Then
\bea
G_0^{(11)}(k;\alpha)=G_0(k)+v^2(k;\alpha)[2\pi i\delta(k^2-m^2)],
\eea
where
\bea
v^2(k;\alpha)=\sum_{s=1}^d\sum_{\lbrace\sigma_s\rbrace}2^{s-1}\sum_{l_{\sigma_1},...,l_{\sigma_s}=1}^\infty(-\eta)^{s+\sum_{r=1}^sl_{\sigma_r}}\,\exp\left[{-\sum_{j=1}^s\alpha_{\sigma_j} l_{\sigma_j} k^{\sigma_j}}\right],\label{BT}
\eea
is the generalized Bogoliubov transformation \cite{GBT}, with $d$ being the number of compactified dimensions, $\eta=1(-1)$ for fermions (bosons) and $\lbrace\sigma_s\rbrace$ denotes the set of all combinations with $s$ elements.

An important note, in phase space the Green function is dependent of the parameters $q$ and $p$, i.e., $G_0^{(ab)}(x-x';\alpha)=G_0^{(ab)}(q^\mu- q'^\mu,  p^\mu - p'^\mu;\alpha)$.

\section{Energy-momentum tensor for the scalar field in phase space}\label{em}

The lagrangian that describes the scalar field in phase space is given by
\begin{eqnarray}
 \mathcal{L}_{\phi}
 &=&-\frac{1}{4}\frac{\partial  \phi }{ \partial q_\mu}
 \frac{\partial  \phi^{\ast}}{ \partial q^\mu}
+\frac{1}{2}i p^\mu
\left(\phi^{\ast}\frac{\partial\phi }{\partial q^\mu}
-\phi \frac{\partial \phi^{\ast}}{\partial q^\mu}
 \right)-
(p^\mu p_\mu-m^2)\phi \phi^{\ast}.
\end{eqnarray}
In order to calculate the Casimir effect we need the energy-momentum tensor that is defined as
\bea
 \mathcal{T^{\mu \nu}}&=&\frac{\partial {\cal L}_\phi}{\partial(\partial_\mu \phi)}\partial^\nu \phi-g^{\mu\nu}{\cal L}_\phi,\nonumber\\
 &=& - \frac{1}{4} \left(  \frac{\partial  \phi^{\ast}}{ \partial q_\mu}  \frac{\partial  \phi}{  \partial q_\nu} +
  \frac{\partial  \phi}{ \partial q_\mu}  \frac{\partial  \phi^{\ast}}{  \partial q_\nu} \right)
 + \frac{1}{2}i p^\mu  \left( \phi^{\ast} \frac{\partial  \phi}{  \partial q_\nu} -
 \phi \frac{\partial \phi^{\ast}}{  \partial q_\nu} \right) \notag\\
 &-& g^{\mu\nu} \left[
 - \frac{1}{4}  \frac{\partial  \phi}{ \partial q^\lambda}  \frac{\partial  \phi^{\ast}}{  \partial q_\lambda}
 + \frac{1}{2}i p^\lambda \left( \phi^{\ast} \frac{\partial  \phi}{  \partial q^\lambda} -
 \phi \frac{\partial \phi^{\ast}}{  \partial q^\lambda} \right)
 -(p^\lambda p_\lambda-m^2)\phi \phi^{\ast} \right],
\eea
To avoid divergences, the energy-momentum tensor is written at different space-time points as
\bea
 \mathcal{T^{\mu \nu}}&=&\lim_{q'^\mu\rightarrow q^\mu}\tau\Biggl\{ - \frac{1}{4} \left(  \frac{\partial  \phi^{'\ast}}{ \partial q'_\mu}  \frac{\partial  \phi}{  \partial q_\nu} +
  \frac{\partial  \phi}{ \partial q_\mu}  \frac{\partial  \phi^{'\ast}}{  \partial q'_\nu} \right)
 + \frac{1}{2}i p^\mu  \left( \phi^{'\ast} \frac{\partial  \phi}{  \partial q_\nu} -
 \phi \frac{\partial \phi^{'\ast}}{  \partial q'_\nu} \right) \notag\\
 &-& g^{\mu\nu} \left[
 - \frac{1}{4}  \frac{\partial  \phi}{ \partial q^\lambda}  \frac{\partial  \phi^{'\ast}}{  \partial q'_\lambda}
 + \frac{1}{2}i p^\lambda \left( \phi^{'\ast} \frac{\partial  \phi}{  \partial q^\lambda} -
 \phi \frac{\partial \phi^{'\ast}}{  \partial q'^\lambda} \right)
 -(p^\lambda p_\lambda-m^2)\phi \phi^{'\ast} \right]\Biggl\}\nonumber\\
 &=&\lim_{q'^\mu\rightarrow q^\mu}\left\{\, \Gamma^{\mu\nu}\tau[\phi(q)\phi^{'\ast}(q')]\right\},
\eea
where $\tau$ is the ordering operator and
\bea
\Gamma^{\mu\nu}&=&- \frac{1}{4} \left(  \frac{\partial }{ \partial q'_\mu}  \frac{\partial }{  \partial q_\nu} +
  \frac{\partial }{ \partial q_\mu}  \frac{\partial }{  \partial q'_\nu} \right)
 + \frac{1}{2}i p^\mu  \left(  \frac{\partial  }{  \partial q_\nu} -
 \frac{\partial }{  \partial q'_\nu} \right) \notag\\
 &-& g^{\mu\nu} \left[
 - \frac{1}{4}  \frac{\partial }{ \partial q^\lambda}  \frac{\partial}{  \partial q'_\lambda}
 + \frac{1}{2}i p^\lambda \left( \frac{\partial }{  \partial q^\lambda} -
 \frac{\partial }{  \partial q'^\lambda} \right)
 -(p^\lambda p_\lambda-m^2) \right].
\eea
The vacuum expectation value of the energy-momentum tensor is
\bea
\langle \mathcal{T^{\mu \nu}}(x)\rangle=\lim_{q'^\mu\rightarrow q^\mu}\left\{\, \Gamma^{\mu\nu}\langle 0|\tau[\phi(q)\phi^{'\ast}(q')]|0\rangle\right\}.
\eea
The scalar field propagator in phase space is defined as
\bea
G_0(q^\mu- q'^\mu,  p^\mu - p'^\mu)  &=& \langle 0|\tau[\phi(q)\phi^{'\ast}(q')]|0\rangle\nonumber\\
&=&\int \frac{d^4k}{(2\pi)^4}\frac{ \delta (p^\mu-p^{'\mu}) e^{i \kappa_\mu(q^\mu-q^{'\mu})}}
{\left(\frac{1}{4} \kappa^2 -i\kappa_\mu (p^\mu-p^{'\mu})+(p^\mu-p^{'\mu})+ m^2\right)}.
\eea
Using the identity
\bea
\frac{1}{\chi^\mu\chi_\mu+M^2}=\int_0^\infty e^{-t(\chi^\mu\chi_\mu-M^2)}dt
\eea
with $M^2=m^2-2(p^\mu-p'^\mu)(p^\mu-p'^\mu)$, the propagator becomes
\bea
G_0(q^\mu- q'^\mu,  p^\mu - p'^\mu)
&=& \frac{64M^2}{\pi^2}   \delta (p^\mu - p'^\mu)  e^{-2(p^\mu - p'^\mu)(q^\mu - q'^\mu) }
\notag\\
&\times&  \left( \frac{M \mid q- q' \mid}{2}\right)^{-1}
  \kappa_1 \left( \frac{M \mid q- q' \mid}{2}\right),
\eea
where $\kappa_\nu(z)$ is the Bessel function. Using the doublet notation, the physical energy-momentum tensor in terms of the $\alpha$-parameter is
\bea
 \mathbb{T}^{\mu \nu (ab)}(q)=\lim_{q'^\mu\rightarrow q^\mu}\left\{\, \Gamma^{\mu\nu}\overline{G}_0^{ab}(q^\mu- q'^\mu,  p^\mu - p'^\mu;\alpha)\right\},
\eea
where $\mathbb{T}^{\mu\nu (ab)}(q;\alpha)=\langle {\cal T}^{\mu\nu(ab)}(q;\alpha)\rangle-\langle {\cal T}^{\mu\nu(ab)}(q)\rangle$ and
\bea
\overline{G}_0^{(ab)}(q^\mu- q'^\mu,  p^\mu - p'^\mu;\alpha)=G_0^{(ab)}(q^\mu- q'^\mu,  p^\mu - p'^\mu;\alpha)-G_0^{(ab)}(q^\mu- q'^\mu,  p^\mu - p'^\mu).
\eea

\section{Some Applications}\label{applications}

In this sections the Stefan-Boltzmann law and the Casimir effect at finite temperature are calculated.

\subsection{Stefan-Boltzmann law}

Here the $\alpha$ parameter is $\alpha=(\beta,0,0,0)$ and the generalized Bogoliubov transformation takes the form
\bea
v^2(\beta)=\sum_{l_0=1}^{\infty}e^{-\beta k^0l_0}\label{BT1}
\eea
and the Green function becomes
\bea
\overline{G}_0^{(ab)}(q^\mu- q'^\mu,  p^\mu - p'^\mu;\beta)=2\sum_{l_0=1}^{\infty}G_0(q^\mu- q'^\mu-i\beta l_0n_0,p^\mu - p'^\mu),\label{GF1}
\eea
where $n_0=(1,0,0,0)$, is a time-like vector.  For $\mu=\nu=0$, the energy-momentum tensor becomes
\bea
 \mathbb{T}^{00 (11)}(\beta)&=&2\lim_{q'^\mu\rightarrow q^\mu}\left\{\,\sum_{l_0=1}^{\infty} \Gamma^{00}G_0(q^\mu- q'^\mu-i\beta l_0n_0,p^\mu - p'^\mu)\right\}\nonumber\\
 &=& \sum_{l_0} \frac{128}{\pi^2} M^2  \delta (p^\mu - p'^\mu) e^{2i(p_0 - p'_0)(\beta l_0) }
 \bigg\{
 \kappa_1 \left( \frac{M}{2}  il_0\beta
 \right)
   \notag\\
&\times&
\left( p^\mu p_\mu+M^2 \right)
\left(\frac{M}{2} il_0\beta \right)^{-1}- \frac{1}{8 l_0^2 \beta^2} \bigg[ 12\,\kappa_0 \left( \frac{M}{2}  il_0\beta
 \right) \nonumber\\
 &+& \frac{1}{iMl_0\beta}(24- i^2l_0^2 M^2 \beta^2  )
  \,\kappa_1 \left( \frac{M}{2}  il_0\beta
 \right)\bigg] \bigg\}.
\eea
This is the Stefan-Boltzmann law in phase space. This result becomes $\mathbb{T}^{00 (11)}(\beta)\sim T^4$, when it is projected for the usual quantum mechanics space.

\subsection{Casimir effect at zero temperature}

Now  $\alpha=(0,0,0,i2d)$ then
\bea
v^2(d)=\sum_{l_3=1}^{\infty}e^{-i2d k^3l_3}\label{BT2}
\eea
and the Green function is
\bea
\overline{G}_0(q^\mu- q'^\mu,  p^\mu - p'^\mu ;d)=2\sum_{l_3=1}^{\infty}G_0(q^\mu- q'^\mu-2d l_3n_3,  p^\mu - p'^\mu)\label{GF2}
\eea
is the Green function with $n_3=(0,0,0,1)$, being the space-like vector. The energy-momentum tensor for this case becomes
\bea
 \mathbb{T}^{33 (11)}(\beta)&=&2\lim_{q'^\mu\rightarrow q^\mu}\left\{\,\sum_{l_3=1}^{\infty} \Gamma^{33}G_0(q^\mu- q'^\mu- 2d l_3n_3,p^\mu - p'^\mu)\right\}\nonumber\\
 &=& \sum_{l_3 }\frac{128}{\pi^2} M^2  \delta (p^\mu - p'^\mu)  e^{2i(p_3 - p'_3)dl_3 }
   \notag\\
&\times & \bigg\{
 \kappa_1 \left( \frac{M}{2}  idl_3
 \right)
+\left(  - p^\mu p_\mu+M^2 \right)
\left(\frac{M}{2} idl_3 \right)^{-1}
   \notag\\
 &+& \frac{1}{16 l_3^2 d^2} \bigg[ 6\kappa_0 \left( d l_3 M \right)
 + \frac{(12 + d^2l_3^2 M^2 )}{d l_3 M}
  \kappa_1 \left( d l_3 M  \right)\bigg] \bigg\}.
 \eea
 It is the Casimir pressure at zero temperature in phase space. In the standard quantum mechanics space the usual result is recovered.

 \subsection{Casimir effect at finite temperature}

 In this case the $\alpha$ parameter is choice as $\alpha=(\beta,0,0,i2d)$. The generalized Bogoliubov transformation is given by
 \bea
v^2(\beta,d)&=&\sum_{l_0=1}^\infty e^{-\beta k^0l_0}+\sum_{l_3=1}^\infty e^{-i2dk^3l_3}+2\sum_{l_0,l_3=1}^\infty e^{-\beta k^0l_0-i2dk^3l_3}.\label{BT3}
\eea
The first two terms are associated with the Stefan-Boltzmann law and the Casimir effect at zero temperature. The Green function for the third term in eq. (\ref{BT3}), our interest here, is
\bea
\overline{G}_0(q^\mu- q'^\mu,  p^\mu - p'^\mu;\beta,d)=4\sum_{l_0,l_3=1}^\infty G_0\left(q^\mu- q'^\mu-i\beta l_0n_0-2dl_3n_3, p^\mu - p'^\mu\right).\label{GF3}
\eea
Then the Casimir pressure at finite temperature in phase space is given as
\bea
\mathbb{T}^{33 (11)}(\beta,d)&=&4\lim_{q'^\mu\rightarrow q^\mu}\left\{\,\sum_{l_0,l_3=1}^{\infty} \Gamma^{33}G_0(q^\mu- q'^\mu-i\beta l_0n_0- 2d l_3n_3,p^\mu - p'^\mu)\right\}\nonumber\\
&=&  \sum_{l_0,l_3} \frac{256 M^2 }{\pi^2}   \delta (p^\mu - p'^\mu)  e^{2i(p_3 - p'_3)dl_3 }
   e^{2i(p_0 - p'_0)(\beta l_0) }\\
&\times & \bigg\{
 \kappa_1 \left[ \frac{M}{2} \sqrt{ (2 d l_3)^2 + (l_0 \beta)^2}
 \right]+\left( - p^\mu p_\mu+M^2 \right)
\left[\frac{M}{2} \sqrt{ (2 d l_3)^2 + (l_0 \beta)^2}\right]^{-1}\nonumber\\
& +& \frac{1}{2} \bigg[ \frac{(3 (2dl_3)^2-(l_0 \beta)^2 ) }{((2 d l_3)^2 + (l_0 \beta)^2 )^2}
 \kappa_0 \left( \frac{M}{2} \sqrt{(2 d l_3)^2 + (l_0 \beta)^2}
 \right) \notag\\
 &+&
 \frac{(48 (2dl_3)^2+ (2 dl_3)^4 M^2  -(l_0 \beta)^2 (16+ (l_0 M \beta)^2 ))}{M(
  (2 d l_3)^2 + (l_0 \beta)^2)^{\frac{5}{2}}}
  \kappa_1 \left( \frac{M}{2} \sqrt{ (2 d l_3)^2 + (l_0 \beta)^2}
 \right)\bigg] \bigg\}.\nonumber
\eea
This result contain the effect of both time and space compactification. In the same way, in the usual quantum mechanics space the standard result is recovered.

\section{Conclusions}

In this article the introduction of temperature into phase space was explored. We defined the scalar field in phase space by means the use of invariants of the respective relativistic algebra. Then we presented the energy-momentum tensor in such a space. This result was used to implement the prescription of Thermo Field Dynamics which allows to deal with some phenomena at finite temperature such as the analogous Casimir effect and Stefan-Boltzmann law. We point out that we projected the mean energy and pressure in the space of coordinates in order to recover the results of literature. If we project our result in the momenta space we should obtain a fundamental energy associated to the given temperature. Such a result should be better understood since the existence of this thermal energy affects the interpretation of phase space.


\begin{thebibliography}{99}
\bibitem{Wigner} E. Wigner, Phys. Rev. \textbf{40}, 749 (1932).
\bibitem{Hillery} M. Hillery, R. F. O'Connel, M. O. Scully, E. P. Wigner, Phys. Rep. \textbf{106}, 121 (1984).
\bibitem{Oliveira} M.D. Oliveira, M.C.B. Fernandes, F.C. Khanna,A.E. Santana, J.D.M.Vianna,  Ann. Phy. \textbf{312},  492  (2004).
\bibitem{Ronni} R.G.G. Amorim , M.C.B. Fernandes , F.C. Khanna , A.E. Santana, J.D.M. Vianna,  Phy. Lett. A \textbf{361}, 464  (2007).
\bibitem{Ronni01} R. G. G. Amorim, F. C. Khanna, A. P. C. Malbouisson, J. M. C. Malbouisson, A. E. Santana,
Int. J. Mod. Phys.A \textbf{30}, 1550135 (2015).
%
\bibitem{wick} T. Matsubara,  Prog. Theor. Phys. \textbf{14}, 351 (1955).
%
\bibitem{khanna} F. C. Khanna, A. P. C. Malbouisson, J. M. C. Malboiusson and A. E. Santana, Themal quantum field theory: Algebraic aspects and applications, World Scientific, Singapore, (2009).

\bibitem{Casimir} H. G. B. Casimir, Proc. K. Ned. Akad. Wet. {\bf 51}, 793 (1948).
\bibitem{artigokhanna} S. C. Ulhoa, A. F. Santos and Faqir C. Khanna, Int. J. Mod. Phys. A {\bf 32}, 1750094 (2017).  
\bibitem{Sparnaay} M. J. Sparnaay, Physica {\bf 24}, 751 (1958).
\bibitem{Lamoreaux} S. K. Lamoreaux, Phys. Rev. Lett. {\bf 28}, 5 (1997).
\bibitem{Mohideen} U. Mohideen and A. Roy, Phys. Rev. Lett. {\bf 81}, 21 (
%
\bibitem{amorim2} R.G.G. Amorim, F.C. Khanna, A.E. Santana, J.D.M. Vianna,
Physica A \textbf{388}, 3771 (2009).
%
%
\bibitem{Umezawa1}Y. Takahashi and H. Umezawa, Coll. Phenomena \textbf{2}, 55 (1975); Int. Jour. Mod. Phys. B \textbf{10}, 1755 (1996).

\bibitem{Umezawa2}Y. Takahashi, H. Umezawa and H. Matsumoto, Thermofield Dynamics and Condensed States, North-Holland, Amsterdan, (1982).
%\bibitem{Umezawa2}Y. Takahashi, H. Umezawa and H. Matsumoto, Thermofield Dynamics and Condensed States, North-Holland, Amsterdan, (1982); F. C. Khanna, A. P. C. Malbouisson, J. M. C. Malboiusson and A. E. Santana, Themal quantum field theory: Algebraic aspects and applications, World Scientific, Singapore, (2009).
\bibitem{Umezawa22} H. Umezawa, Advanced Field Theory: Micro, Macro and Thermal Physics, AIP, New York, (1993).
\bibitem{Khanna1} A. E. Santana and F. C. Khanna, Phys. Lett. A \textbf{203}, 68 (1995).
\bibitem{Khanna2} A. E. Santana, F. C. Khanna, H. Chu, and C. Chang, Ann. Phys. \textbf{249}, 481 (1996).
\bibitem{GBT}F. C. Khanna, A. P. C Malbouisson, J. M. C. Malbouisson and A. E. Santana, Ann. Phys. \textbf{326}, 2634 (2011).
%
%
\bibitem{amorim4} R.G.G. Amorim, M.C.B. Fernandes, F.C. Khanna, A.E. Santana,
J.D.M. Vianna, Int. J. Mod. Phys. A \textbf{28}, 1350013 (2013).

\bibitem{amorim5} R.G.G. Amorim, S.C. Ulhoa, A.E. Santana, Braz. J. Phys.
\textbf{43}, 78 (2013).


\bibitem{wig3} Y. S. Kim and M. E. Noz, \textit{Phase Space Picture and
Quantum Mechanics - Group Theoretical Approach (}W. Scientific,
London, 1991).

\bibitem{wig4} T. Curtright, D. Fairlie and C. Zachos, Phys. Rev. D \textbf{58}, 25002 (1998).

\bibitem{gal1} D. Galetti and A. F. R. de Toledo Piza, Physica A \textbf{214}, 207 (1995).

\bibitem{hor1} L. P. Horwitz, S. Shashoua and W. C. Schive, Physica A
\textbf{161}, 300 (1989).

\bibitem{hak3} P. R. Holland, Found. Phys. \textbf{16}, 701 (1986).
\bibitem{bos} M. A. de Gosson, J. Phys. A:
Math. Gen. \textbf{38}, 1 (2000).

\bibitem{wig55} T. Curtright and C. Zachos, J. Phys. A \textbf{32}, 771 (1999).
%

\end{thebibliography}
\end{document}